# Interference Effects Due to Commensurate Electron Trajectories and Topological Crossovers in (TMTSF)$_2$ClO$_4$

H.I. Ha[1,*], A.G. Lebed[2,†] and M.J. Naughton[1]

[1]Department of Physics, Boston College, Chestnut Hill, Massachusetts 02467, USA

[2]Department of Physics, University of Arizona, Tucson, Arizona 85721, USA

**Abstract**

We report angle-dependent magnetoresistance measurements on (TMTSF)$_2$ClO$_4$ that provide strong support for a new macroscopic quantum phenomenon, the "interference commensurate" (IC) effect, in quasi-one dimensional metals. In addition to observing rich magnetoresistance oscillations, and fitting them with one-electron calculations, we observe a clear demarcation of field-dependent behavior at local resistance minima and maxima (versus field angle). Anticipated by a theoretical treatment of the IC effect in terms of Bragg reflections in the extended Brillouin zone, this behavior results from 1D → 2D topological crossovers of electron wave functions as a function of field orientation.

In recent years, several new types of angular magnetoresistance oscillation (AMRO) phenomena have been discovered in quasi-one dimensional metals such as the organic conductors (TMTSF)$_2$X [1]. Such materials can be envisioned by considering one's hand. The fingers represent highly conducting chains along *x* that are weakly coupled within the hand along *y*, forming a conducting layer with in-plane anisotropy. The hands are themselves coupled even more weakly along *z*, forming a highly anisotropic, 3D system. The AMRO effects occur when a magnetic field is rotated within different planes in this system.

For example, *y-z* plane field rotations in (TMTSF)$_2$ClO$_4$ [2,3,4,5] and (TMTSF)$_2$PF$_6$ [6,7] reveal the Lebed [8] "magic angle" effect, wherein the interlayer (interhand) resistivity develops pronounced minima at characteristic orientations given by $N(b/c)\tan\theta = 1$, where *N* is an integer, *b*//*y* and *c*//*z* are interchain (interfinger) and interplane (interhand) crystal spacings, respectively, and $\theta$ is the tilt angle from *b* to *c*. Resonant oscillations up to ~10$^{th}$ order in *N* have been observed to date. As first shown by Danner, Kang and Chaikin [9], rotating in the *x-z* plane results in the appearance of a single resistivity peak centered about a small tilt angle from B//*x* (from the fingers). Finally, rotations from *x* to *y* (within the plane of the hand) result in the "third angular effect" (TAE) which is manifest as a resistance minimum near B//*x* [10,11,12].

There have been many proposed explanations of the *y-z* plane magic angle effect, [8,13,14,15,16] including some involving interactions between electrons [7,8,15]. While these theories share many common aspects, a consensus is still lacking. There even exists experimental evidence [7] of a non-Fermi liquid origin to some magic angle effects. The *x-z* plane effect is simpler to understand, being tied directly to details of the warped open Fermi surface. Through this effect, the transfer integral ratio $t_b/t_a$ may be measured [9]. Since $t_a \sim E_F$, the Fermi energy, and can be independently determined, this effect provides a means to measure $t_b$, the interchain coupling, or the degree to which the Fermi sheets are warped in the $k_b$ direction. The *x-y* plane TAE is ascribed to the velocity-preserving nature of "effective" electrons, via their proximity to geometrical inflection points on the Fermi surface [17]. As such, this can be used to quantify the interplane coupling $t_c$ as well as $t_b$. These latter two AMRO effects result, to varying degrees, from the semiclassical behavior of electrons on a warped, open Fermi surface in a magnetic field, and can be explained by solving the Boltzmann kinetic equation in one Brillouin zone [9,17].

When one rotates a magnetic field through something other than one of these principal planes, the simplest expectation is for some trivial combination of the above three AMRO effects. In fact, in the first such experiment [12] on (TMTSF)$_2$PF$_6$, *x-y* plane rotations with a finite field component along *z*, a rich AMRO spectrum was observed. These so-called "LN oscillations" were first interpreted in the simple manner above, with resistivity minima ascribed to projections of commensurate directions according to $N(b/c)\tan\theta = \sin\phi$ (N = integer and $\phi$ is the tilt angle from *a* to *b*) [12]. However, it has recently been suggested via analytical theory that the origin of these new complex oscillations is instead related to interference effects due to commensurate electron trajectories in a tilted magnetic field in the extended Brillouin zone [18,19]. Moreover, as shown in Ref. [19], the interference commensurate (IC) effect leads to topological 1D → 2D crossovers at commensurate directions of a magnetic field. In this Letter, we provide evidence for this IC interpretation in the sister compound (TMTSF)$_2$ClO$_4$ via angle-dependent magnetoresistance measurements. We use the IC theory to accurately calculate the angle dependence, and we report a new manifestation of the IC effect in fixed-angle experiments, where the magnetoresistance displays qualitatively different behavior at field orientations when the resistance is at a local maximum versus a minimum, corresponding to 1D and 2D transport, respectively.

The (TMTSF)$_2$X (X= ClO$_4$, PF$_6$, etc.) family of organic conductors can be modeled more quantitatively by a tight-binding band approximation, with electron band energy given by $E_k = -2t_a \cos(k_a a/2) - 2t_b \cos(k_b b) - 2t_c \cos(k_c c)$.[1] With the strongest overlap of π-molecular orbitals along the *a*-direction (*x*, chain, finger), the electron energy $E_k$ is mostly dominated by $k_a$ and is nearly independent of $k_b$ and $k_c$. Therefore, the Fermi surface consists of a pair of warped sheets, defined by the electron transfer energies $t_a$, $t_b$, and $t_c$ along the three crystal axes. These latter two are the inter-finger and inter-hand couplings, respectively, discussed above. Because of the open Fermi surface, traditional magnetic oscillations [20] cannot exist in these materials. Nonetheless, the metallic phase of (TMTSF)$_2$X exhibits the large number of unconventional magnetic oscillations directly related to this open Q1D Fermi surface discussed above.

The geometry of our experiment is shown in the inset to Figure 1. We show the two angles $\theta$ and $\phi$, corresponding to magnetic field rotation directions out of and within the *x-y* (*a-b*) plane, respectively. Two (TMTSF)$_2$ClO$_4$ crystals were mounted with 12-μm gold wires and graphite paste on an *in situ* $\theta$-rotating stage of a dilution refrigerator, situated in a $\phi$-rotatable split-coil superconducting magnet. The $\phi$-dependence of the interlayer resistivity $\rho_{zz}$ for various angles $\theta$ was measured at fixed temperature and field. The two samples showed nearly identical behavior; results for one are shown here. We show in Fig. 1(a) the magnetoresistance [ρ(*B*)−ρ(0)]/ρ(0) ≡ Δρ/ρ at 0.1K and 10T, where ρ(0) is the extrapolated zero-field resistivity at 0.1 (*i.e.* ignoring the superconducting transition). For clarity, the same data are shown in Fig. 1(b) with vertical offsets. The bottom curve there represents a pure *x-y* or *a-b* plane rotation without any *c*-axis field component ($\theta$=0). The double minimum seen there, corresponding to the TAE, is distinctive up to small tilt angles, $\theta <$ ~3°, and is then slowly smeared out as richer (LN) oscillations develop at larger $\theta$. The positions of these LN oscillation minima correspond to commensurate magnetic field directions [19] $2N(b/c)\tan\theta = \sin\phi$, where the coefficient 2 is due to the reduced size of the Brillouin zone (b→2b) in (TMTSF)$_2$ClO$_4$.

In the IC theory [18,19], the origin of the LN oscillations is related to special commensurate electron trajectories where the average electron velocity along $z$ is non-zero. This analytical theory [18] well describes previous data [12] on (TMTSF)$_2$PF$_6$ whereas, for the present case, a more complex approach is needed to account for the doubling of the unit cell along the b-axis due to ordering of the ClO$_4$ anions. The details of this approach are published elsewhere [19]. The resulting conductivity, accounting for the anion ordering gap $\Delta_{AO}$, is given by:

$$\sigma_{zz} \sim \int_{-\infty}^{0} dt\, e^{t/\tau} \int_{0}^{2\pi} \frac{dy}{2\pi} \cos[\omega_c t + \frac{\omega_c^*}{\omega_b}\left(\sqrt{\cos^2(\omega_b t + y) + A^2} - \sqrt{\cos^2 y + A^2}\right)], \text{ where } \omega_b = ev_F bB$$

$\sin\theta$, $\omega_c = ev_F cB\cos\theta\sin\phi$, $\omega_c^* = ev_y^o cB\cos\theta\cos\phi$, $v_y^o = 2t_b b$, $y = k_y b$, and $A = \Delta_{AO}/2t_b$.

In Fig. 2, we compare the calculated conductivity using this formalism with the data of Fig. 1, using the same values of $t_a/t_b = 9.75$, $\omega_c\tau = 15$ at $B = 10$T and $\Delta_{AO}/2t_b = 0.10$ for all three curves. The first and last quantities were used as fitting parameters, while the middle one was arrived at using a published value of relaxation time $\tau$ [12]. These curves demonstrate not only qualitative but rather good quantitative agreement between theory and experiment over a broad range of magnetic field orientations. (The data are asymmetric with respect to zero angle $\phi$, since the actual crystal structure is triclinic, while the calculations are for an orthorhombic approximation, with only positive angle data chosen for the fitting.)

The above suggests that the interference commensurate phenomenon may be an ideal tool to determine all band parameters in Q1D conductors, such as the in-plane anisotropy $t_a/t_b$ and the anion ordering gap $\Delta_{AO}$, and it can be also used for a determination of the non-trivial topology of a Fermi surface. Unlike $(TMTSF)_2PF_6$, due to the anion ordering in $(TMTSF)_2ClO_4$, not only is the size of the Brillouin zone reduced, but also the shape of Fermi surface is changed. Thus, when magnetic field is applied, the electron trajectories in these two systems are quite different. Especially in the absence of any magnetic breakdown, electrons travel on only one part of the Fermi surface in $(TMTSF)_2ClO_4$. Note that the positions of the LN oscillation minima are defined only by the dimensionality of electron wave functions in the magnetic field and correspond to the condition [19] $2N(b/c)\tan\theta = \sin\phi$ in $(TMTSF)_2ClO_4$. Thus, the shape and amplitudes of the oscillations are affected by the shape of the FS such that, by their analysis, we can define the shape of the FS more accurately. In conventional metals [20] with closed Fermi surfaces, the dHvA effect is traditionally used to obtain the shape of the Fermi surface, but this can't been done on open Fermi surfaces, such as appear in $(TMTSF)_2X$. The present technique provides such capability.

The anion order gap parameter used in the above fits can be compared to a value extracted from magnetic breakdown across the anion gap, corresponding to $\omega_c$ of the order of $\Delta_{AO}^2/t_b$. Using the Gor'kov-Lebed equations from Ref. 21, the probability of an electron orbit experiencing magnetic breakdown varies as $\exp(-\pi\Delta_{AO}^2/2t_b\omega_b)$. Using a breakdown field of 15T from Yan, *et al.* [22] yields $\Delta_{AO}$ = 56 K (for $t_b$=200K and $\omega_c$=1.5 K/T), corresponding to a ratio $\Delta_{AO}/2t_b = 0.14$. This is indeed comparable to the 0.10 ratio used in our fits above, which yielded an anion gap of $\Delta_{AO}$ ~ 40 K, as well as to the value of 0.14 from Yoshino *et al.* [23] (or 0.17, depending on whether one uses our $t_a/t_b$=9.75, or their ratio of 12) from a study of the TAE under pressure. Analysis of these oscillations now provides the most accurate way to determine this anion ordering gap.

The key point of this work, which was not appreciated in our earlier Letter [18], is that the magnetoresistance oscillations in Figs. 1 and 2 can be interpreted in terms of unique 1 to 2–dimensional crossovers [19]. In the absence of Landau quantization for open Fermi surfaces, the "other" quantum effect in a magnetic field, Bragg reflections, results in a series of 1$D$ to 2$D$ crossovers at the LN oscillation minima. In other words, electron wave functions, localized on chains (*i.e.* 1D) at arbitrary field directions [8,15], become delocalized on planes (*i.e.* 2D) at the commensurate directions. The non-trivial physical origin of these 1$D \rightarrow$ 2$D$ crossovers is related to interference effects between velocity components along the *z*-axis, $k_z$, and electron motion along the *y*-direction [19]. These interference effects occur as electrons move along open FS sheets in the extended Brillouin zone and are qualitatively different from that responsible for the magic angle effect [16]. A discussion of how 1$D \rightarrow$ 2$D$ dimensional crossovers can lead to the appearance of oscillation minima in $\rho_{zz}$ can be found in Ref. 19.

We can use this dimensional crossover notion and the IC model to further discuss electron motion in such quasi-1D metals in a strong magnetic field. For electrons localized on conducting *x*-chains, one expects, in the absence of impurities ($1/\tau = 0$), the conductivity between chains $\sigma_{zz}$ to be zero. For $1/\tau \neq 0$, σ is finite and should vary as $1/\omega_c^2\tau^2 \sim 1/B^2$ at high field. If, at commensurate field directions, electron wave functions become delocalized (2$D$), then $\sigma_{zz}$ is expected to be similar to conductivity in the absence of a magnetic field. Thus, $\sigma_{zz}$ should saturate at high field and be proportional to $\tau$. In other words, as a consequence of the 1D$\rightarrow$2D crossovers, $\rho_{zz}(B,\theta,\phi) \equiv 1/\sigma_{zz}$ should saturate at high field for commensurate orientations, and tend toward $B^2$ behavior away from these special directions. This is in contrast to the conventional behavior of magnetoresistance in an open-orbit metal [24]. Fig. 3 presents the results of field sweeps at several of the minima and maxima of our LN oscillation data of Figs. 1 and 2. As predicted, $\Delta\rho_{zz}/\rho(B,\theta,\phi)$ saturates at commensurate directions (minima), while at non-commensurate directions (maxima), it exhibits nonsaturating behavior. As shown in Fig. 3, there is rather good agreement between experiment and the calculated magnetoresistance using the conductivity equation above and the same parameters as were used for the fits in Fig. 2. Thus, the predicted dramatic difference between commensurate 2D and non-commensurate 1D magnetoresistance, with saturating at commensurate angles (minima in angle sweeps) and non-trivial, non-saturating otherwise, is borne out in experiment and calculation.

In summary, Bragg reflections, which occur as electrons move along open Fermi surface sheets in an extended Brillouin zone in the quasi-1D metal (TMTSF)$_2$ClO$_4$, result in interference effects which represent a novel type of macroscopic quantum phenomenon in a magnetic field, of topological origin. Due to these interference effects, the effective space dimensionality of the electron wave function is changed, resulting in large conductivity oscillations, with $\Delta\sigma/\sigma \gg 1$, far more than is observed to result from Landau quantization of electron orbits in conventional metals. We expect that the method suggested in this Letter for determining the topology and shape of open Fermi surfaces will be found useful for other families of low-dimensional conductors (such as (DMET-TSeF)$_2$X, $\kappa$-(ET)$_2$Cu(NCS)$_2$, $\alpha$-(ET)$_2$ MHg(XCN)$_4$, etc.). It may as well be a reliable tool to test for Fermi-liquid behavior in (TMTSF)$_2$X conductors, which is claimed to be broken [7,25] under certain experimental conditions.

This work was supported by NSF Grant DMR-0308973 and INTAS Grant 2001-0791. We would like to acknowledge Harukazu Yoshino for helpful discussions.

* Current address: Department of Physics, Harvard University, Cambridge, Mass.

† Also L.D. Landau Institute for Theoretical Physics, Moscow, Russia

**References**

[1] T. Ishiguro, K. Yamaji, G. Saito, *Organic Superconductors,* 2nd Edition, Springer-Verlag (1998).

[2] M.J. Naughton, O.H. Chung, L.Y. Chiang, and J.S. Brooks, Mat. Res. Soc. Symp. Proc. **173**, 257 (1990).

[3] T. Osada, A. Kawasumi, S. Kagoshima, N. Miura, and G. Saito: Phys. Rev. Lett. **66**, 1512 (1991).

[4] M.J. Naughton, O.H. Chung, M. Chaparala, X. Bu, and P. Coppens, Phys. Rev. Lett. **67**, 3712 (1991).

[5] G.S. Boebinger, G. Montambaux, M.L. Kaplan, R.C. Haddon, S.V. Chichester, and L.Y. Chiang, Phys. Rev. Lett. **64**, 591 (1990).

[6] W. Kang, S. T. Hannahs, and P. M. Chaikin, Phys. Rev. Lett. **69**, 2827 (1992).

[7] W. Wu, I.J. Lee, and P.M. Chaikin, Phys. Rev. Lett. **91**, 056601 (2003).

[8] A.G. Lebed, Pis'ma Zh. Eksp. Teor. Fiz. **43**, 137 (1986) [JETP Lett. **43**, 174 (1986)]; A.G. Lebed and Per Bak, Phys. Rev. Lett. **63**, 1315 (1989).

[9] G.M. Danner, W. Kang and P.M. Chaikin, Phys. Rev. Lett. **72**, 3714 (1994).

[10] H. Yoshino, K. Saito, K. Kikuchi, H. Nishikawa, K. Kobayashi, and I. Ikemoto, J. Phys. Soc. Jpn. **64**, 2307 (1995); H. Yoshino, *et al.*, J. Phys. Soc. Jpn. **65**, 3127 (1996); H. Yoshino, *et al.*, J. Phys. Soc. Jpn. **66**, 2410 (1997).

[11] T. Osada, S. Kagoshima, and N. Miura, Phys. Rev. Lett. **77**, 5261 (1996).

[12] I.J. Lee and M.J. Naughton, Phys. Rev. B **57**, 7423 (1998); M.J. Naughton, I.J. Lee, G.M. Danner and P.M. Chaikin, Synth. Met. **85**, 1481 (1997).

[13] T. Osada, S. Kagoshima, and N. Miura, Phys. Rev. B **46**, 1812 (1992).

[14] K. Maki, Phys. Rev. B **45**, 5111 (1992).

[15] A.G. Lebed: J. Phys. I (France) **4**, 351 (1994) ; J. Phys. I (France) **6**, 1819 (1996).

[16] A.G. Lebed, N.N. Bagmet and M.J. Naughton, Phys. Rev. Lett. **93**, 157006 (2004).

[17] A.G. Lebed and N.N. Bagmet, Phys. Rev. B **55**, R8654 (1997).

[18] A.G. Lebed and M.J. Naughton, Phys. Rev. Lett. **91**, 187003 (2003).

[19] A.G. Lebed, H.I. Ha and M.J. Naughton, cond-mat/0411206, Phys. Rev. B, in print (2005).

[20] A.A. Abrikosov, *Fundamentals of the Theory of Metals* (Elsevier, Amsterdam 1988).

[21] L.P. Gor'kov and A.G. Lebed, Phys. Rev. B **51**, 3285 (1995).

[22] X. Yan, M.J. Naughton, R.V. Chamberlin, L.Y. Chiang, S.Y. Hsu, and P.M. Chaikin, Synth. Met. **27**, B145 (1988).

[23] H. Yoshino, S. Shodai, and K. Murata, Synth. Met. **133-134**, 55 (2003).

[24] A.B. Pippard, *Magnetoresistance in Metals* (Cambridge Univ. Press, Cambridge 1989).

[25] G.M. Danner and P.M. Chaikin, Phys. Rev. Lett. **18**, 4690 (1995).

**Figure Captions**

**Fig. 1** (a) Raw data of interlayer magnetoresistance Δρ/ρ during *a-b*-plane φ-rotations with *c*-axis magnetic field components $\theta$ at 0.1K and 10T. The curves have a common background resistance. (b) Same data, offset for clarity, with angles $\theta$ indicated at right. (Inset) Rotation angles $\theta$ and $\phi$ with respect to the crystal.

**Fig. 2** Angle-dependent magnetoresistance calculated for $\theta = 5°, 7°,$ and $15°$ (solid lines) compared with the experimental data at 0.1K, 10T (open circles). Fitting parameters are the same for all three curves (see text).
.

**Fig. 3** 1D→2D topological crossovers as revealed by Δρ/ρ(*B*) at certain commensurate and non-commensurate orientations. For each angle $\theta$ indicated, $\phi$ was adjusted to reach a resistance maximum or minimum, as represented by the set of arrows in the $\theta = 6°$ data in Fig. 1. For maxima, the magnetoresistance is nonsaturating in field (1D-like), while for minima, it tends to saturation (commensurate, 2D-like). Consistent dependencies occur in the calculated magnetoresistance (ignoring the superconducting transition at low field), as shown in dashed lines atop each experimental curve.

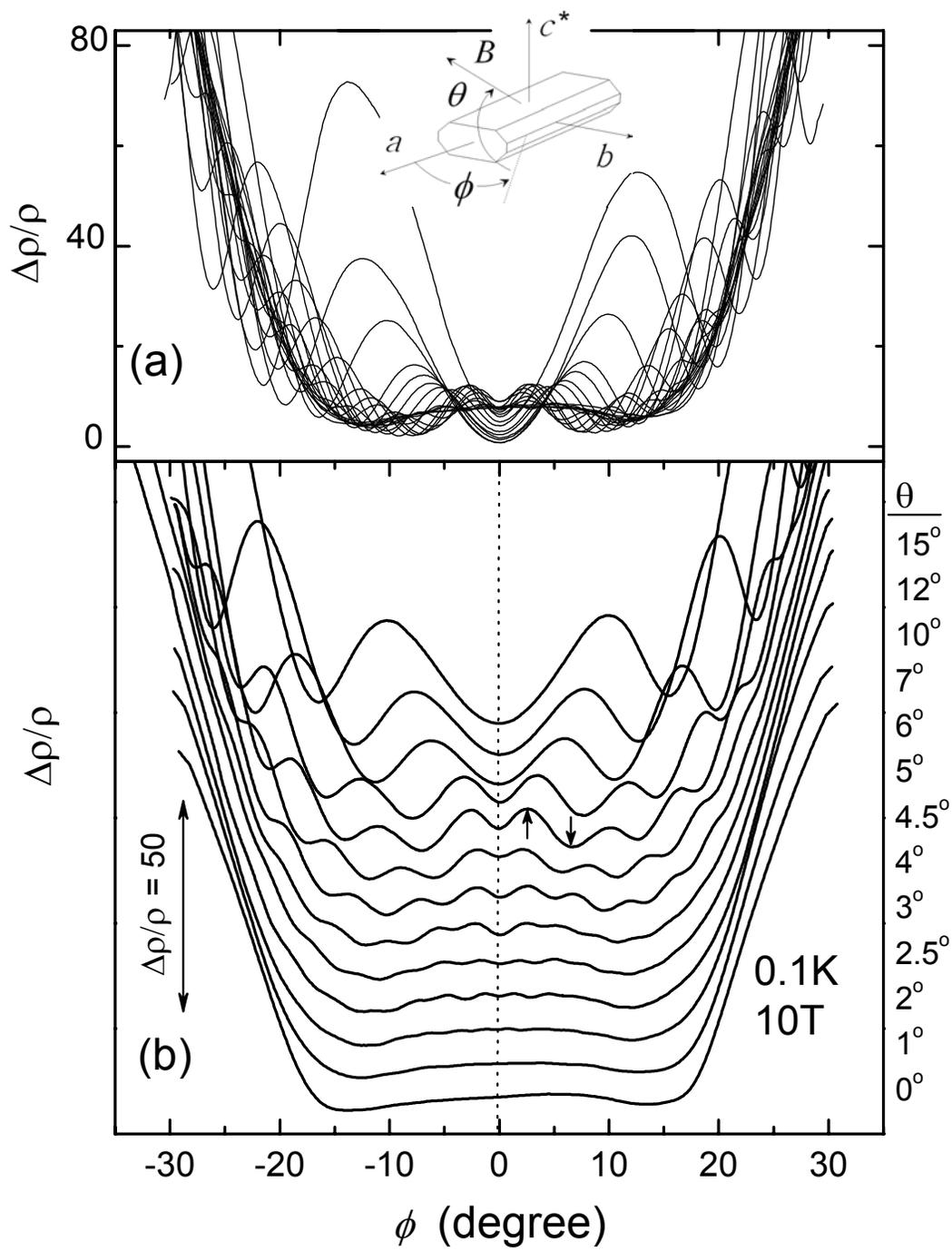

**Fig. 1**

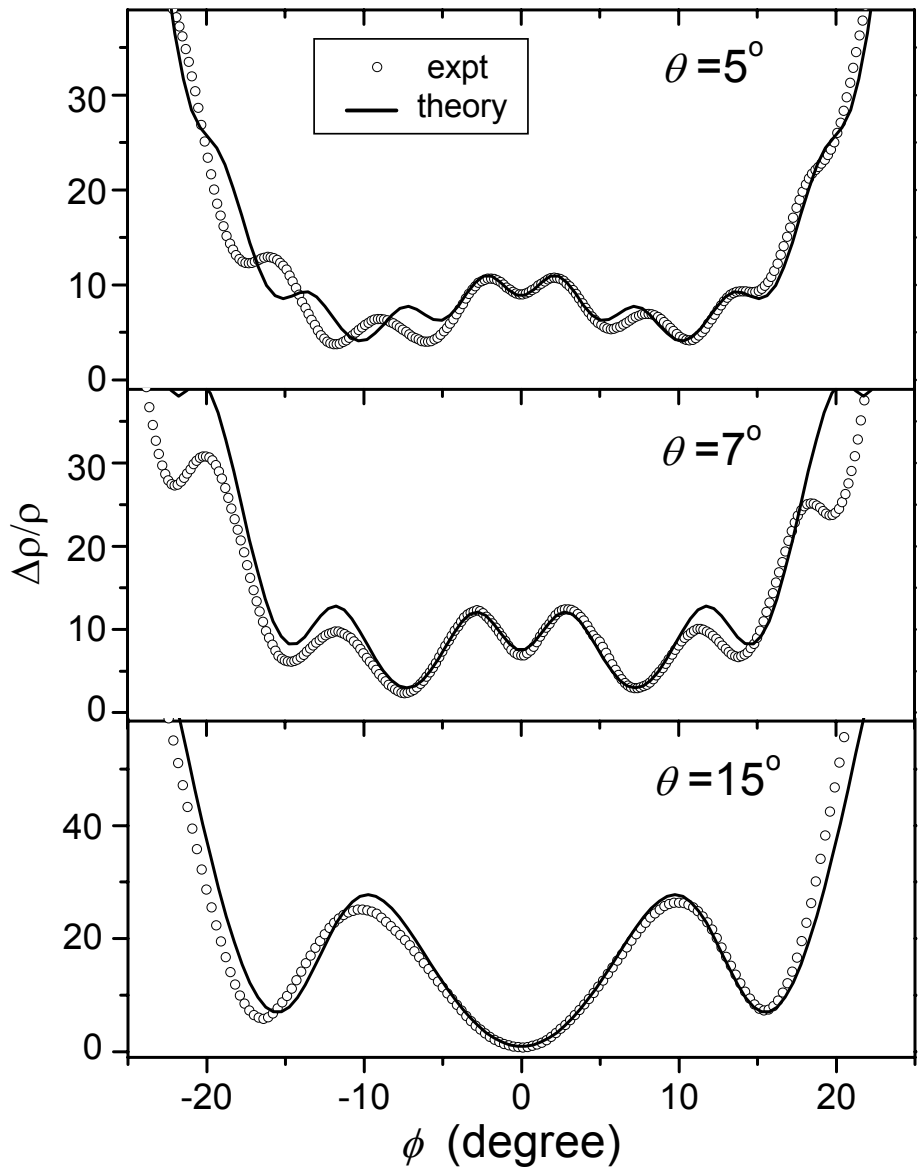

**Fig. 2**

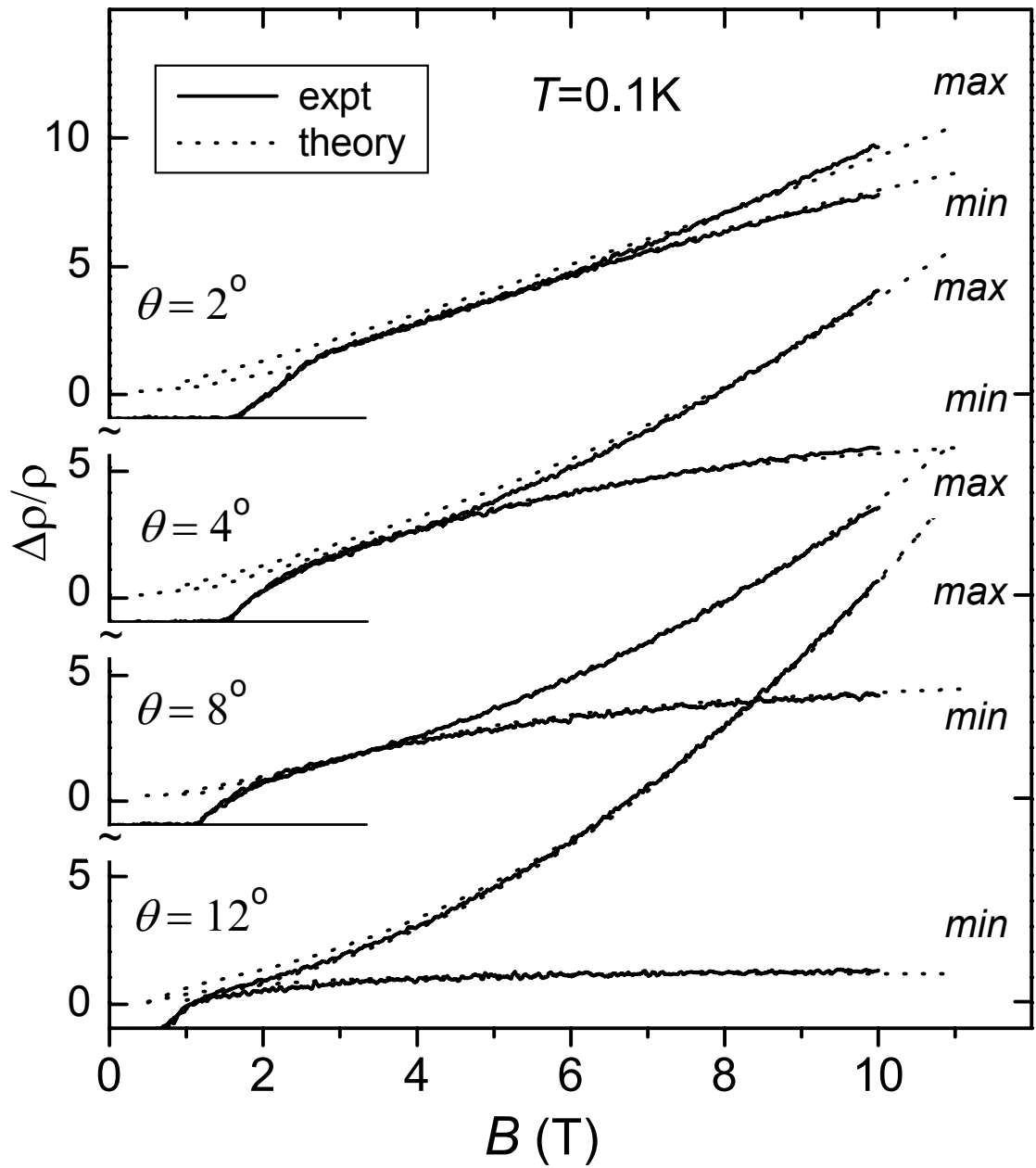

**Fig. 3**